\documentclass[aps,pra,twocolumn,groupedaddress]{revtex4}

\usepackage{graphicx}
\usepackage{amsmath}

\newcommand{\partder}[2]{\frac{\partial{#1}}{\partial{#2}}}

\newcommand{\bal}{\begin{align}}
\newcommand{\eal}{\end{align}}

\newcommand{\be}{\begin{equation}}
\newcommand{\ee}{\end{equation}}
\newcommand{\ba}{\begin{eqnarray}}
\newcommand{\ea}{\end{eqnarray}}
\newcommand{\bastar}{\begin{eqnarray*}}
\newcommand{\eastar}{\end{eqnarray*}}

\begin{document}


\title{Dynamically stable multiply quantized vortices in dilute Bose-Einstein condensates}

\author{J. A. M.~Huhtam\"aki,$^1$ M.~M\"ott\"onen,$^{1,2}$ and S. M. M.~Virtanen$^1$}

\affiliation{$^1$Laboratory of Physics, Helsinki University of
  Technology, POB 4100, FI-02015 TKK, Finland}
\affiliation{$^2$Low Temperature Laboratory, Helsinki University of Technology, POB 3500,
  FI-02015 TKK, Finland,}

\date{\today}

\begin{abstract}
Multiquantum vortices in dilute atomic Bose-Einstein condensates confined in long cigar-shaped traps are known to be both energetically and
dynamically unstable. They tend to split into single-quantum vortices even in the ultralow temperature limit with vanishingly weak
dissipation, which has also been confirmed in the recent experiments~[Y. Shin {\it et al}., Phys. Rev. Lett. \textbf{93}, 160406
(2004)] utilizing the so-called topological phase engineering method to create multiquantum vortices. We study the stability properties of
multiquantum vortices in different trap geometries by solving the Bogoliubov excitation spectra for such states. We find that there are regions
in the trap asymmetry and condensate interaction strength plane in which the splitting instability of multiquantum vortices is suppressed, and
hence they are dynamically stable. For example, the doubly quantized vortex can be made dynamically stable even in spherical traps within a wide
range of interaction strength values. We expect that this suppression of vortex-splitting instability can be experimentally verified.
\end{abstract}

\pacs{PACS number(s): 03.75.Lm, 03.75.Kk}
\keywords{Bose-Einstein condensate, vortex, topological defect}

\maketitle
\section{Introduction}\label{int}
Quantized vortices are fundamental topological excitations in systems having long range quantum phase coherence, {\it e.g.}, in superfluids and
superconductors. Creation and observation of singly quantized vortices in dilute atomic Bose-Einstein condensates (BECs) in
1999~\cite{Matthews1999b} was an important demonstration of the superfluid properties of these gaseous systems. Since then, the properties of
singly quantized vortices and vortex lattices have been studied extensively, both experimentally and theoretically \cite{Fetter2001c}.

However, there has been considerably fewer
investigations of multiply quantized vortices, for
which the phase of the order parameter winds a multiple~$\kappa$ of $2\pi$.
Typically, quantized vortices are generated by rotating the condensate with
a time-dependent external potential. Since the energy of
multiquantum vortices in
harmonically trapped condensates increases considerably
faster than linearly with respect to vorticity quantum number $\kappa$, single-quantum vortices are
favourable when the system is rotated and thus multiquantum vortices do
not appear.

Although difficult to create with the usual rotational techniques, Leanhardt {\it et al.}\ were able to create the first two and four times
quantized vortices in dilute BECs~\cite{Leanhardt2002a} using the so-called topological phase engineering method proposed by Nakahara {\it et
al.}~\cite{Nakahara2000a,Isoshima2000a,Ogawa2002,Mottonen2002}. In this technique, the bias field of the Ioffe-Pritchard trap is reversed almost
adiabatically, which transfers the atoms coherently from the original irrotational spin state to other spin states that contain multiquantum
vortices. In the vicinity of the middle point of the field reversal period the bias field is weak, and hence spin flips may occur close to the
center of the trap. Due to these Majorana flips approximately half of the atoms are transferred into nonconfined states and escape from the
trap. If the total spin quantum number of the condensate atoms is $F=1$, the bias field reversal transfers atoms into the single weak field
seeking state (WFSS) which contains a doubly quantized vortex. On the contrary, if the condensate consists of atoms with $F=2$, there are two
WFSSs. In the theoretical simulations~\cite{Mottonen2002}, both of these WFSSs are macroscopically populated due to Majorana flips after the
field reversal. One of the WFSSs contains a three times quantized vortex and the other a four times quantized one. In the experiments however,
the measured spatial angular momentum per particle was~$(4.4\pm0.4)\hbar$, seeming to match with the angular momentum of a pure four times
quantized vortex~\cite{Leanhardt2002a}. Since no tomographic images of different spin components for $F=2$ condensates after field reversal has
been reported as for the $F=1$ condensates~\cite{Leanhardt2003a}, it still remains somewhat open whether the created state was actually a
mixture of three- and four-quantum vortices.

Theoretical investigations~\cite{Mottonen2003a,Huhtamaki2006a} have shown that multiply quantized vortices are dynamically unstable in the
experimentally realized cigar-shaped geometries~\cite{Leanhardt2002a,Leanhardt2003a,Shin2004a}. This implies that they do not only decay at
finite temperatures in the presence of dissipation, but also small perturbations to a stationary multiquantum vortex in a pure condensate result
in its splitting into $\kappa$ singly quantized vortices. The splitting instability of doubly quantized vortices has also been observed in the
experiments~\cite{Shin2004a}, in which their lifetimes were measured as functions of the particle density. It has also been theoretically
confirmed that the gravitational sag and the time-dependent magnetic potential used in the experiments alone suffice to act as the perturbation
required to initiate the splitting process and lead to agreement with the measured lifetimes even when dissipation is
neglected~\cite{Huhtamaki2006a}. Furthermore, compensation of the gravitational sag with an optical potential has resulted in longer lifetimes
of condensates with four-quantum vortices~\cite{Kumakura2006a}.

To increase the lifetime of multiquantum vortices, one could try to minimize perturbations breaking the rotational symmetry of the vortex state,
for example by arranging the gravitation to be parallel to the symmetry axis of the vortex, or try to choose the system parameters such that the
state possibly becomes dynamically stable. Suppression of the rotational perturbations is a difficult way to increase the lifetime
significantly, since the amplitudes of the splitting instability modes begin to grow exponentially in time. In the latter possibility, the vortex state
would not decay in the temporal evolution under small perturbations in time
scales for which dissipative effects can be neglected.

In this paper, we show that two-, \mbox{three-,} and four-quantum vortices in harmonically trapped condensates can be made fully dynamically stable by
tuning the interaction strength of the condensate and the aspect ratio of the atom trap suitably. It is known that in the limit of effectively
two-dimensional pancake-shaped condensates there exist dynamical stability regions with respect to the interaction
strength~\cite{Pu1999a,Mottonen2003a,Kawaguchi2004}. Our results show that these stability regions extend nontrivially to genuine three
dimensional geometries, and for example the doubly quantized vortex has wide stability regions as a function of the interaction strength even
for spherical and slightly prolate condensates. The stability regions of three- and four-quantum vortices are smaller, but we show that also
their splitting instabilities can be suppressed for oblate and spherical condensates. Furthermore, since the interaction strength parameter can
be easily tuned by changing the condensate particle number or by utilizing atomic Feshbach resonances, we expect that this stabilization of
multiquantum vortices can be verified experimentally.

\section{\label{equ}Methods}
At ultralow temperatures, the complex valued order parameter field $\psi({\bf r})$ of
a weakly interacting BEC is described
by the Gross-Pitaevskii (GP) equation
\begin{eqnarray}
i\hbar \partder{}{t} \psi({\bf r}, t)=\left[\mathcal{H}+g|\psi({\bf r}, t)|^2
\right]\psi({\bf r},t),
\end{eqnarray}
where $\mathcal{H}=-\frac{\hbar^2}{2m}\nabla^2 + V_{\textrm{ext}}({\bf r})$,
$m$ is the mass of the confined atoms and $V_{\textrm{ext}}({\bf r})$
the external potential. The interaction strength parameter $g$ is related
to the vacuum $s$-wave scattering length by $g=4\pi \hbar^2 a/m$.
Throughout this paper we consider condensates confined in rotationally
symmetric harmonic traps of the form
\begin{eqnarray}\label{trap}
V_{\textrm{ext}}=\frac{1}{2}m\omega_r^2 (r^2+\lambda^2 z^2),
\end{eqnarray}
where $r^2=x^2+y^2$ and $\lambda=\omega_z/\omega_r$ is the ratio between the trapping frequencies in the axial and radial directions. The order
parameter is normalized according to $\int |\psi({\bf r})|^2 \textrm{d}{\bf r}=N$, where $N$ is the total particle number.

Stationary states of the system with an eigenvalue~$\mu$ are of the form $\psi({\bf r})e^{-i\mu t/\hbar}$. Small-amplitude oscillations about a
given stationary state $\psi({\bf r})e^{-i\mu t/\hbar}$ can be studied by writing the order parameter as
\begin{eqnarray}\label{statdec}\psi({\bf r},t)=\left[ \psi({\bf r})+\vartheta({\bf r},t) \right]e^{-i\mu t/\hbar}.\end{eqnarray} By substituting
this trial function into the GP equation with the decomposition \begin{eqnarray}\label{vartheta}\vartheta({\bf r},t)=\sum_q [u_q({\bf r})e^{-i\omega_q
t}+v_q^*({\bf r})e^{i\omega_q t}]\end{eqnarray} and linearizing with respect to $\vartheta({\bf r},t)$, we obtain the Bogoliubov equations
\begin{equation}\label{bogo}
\begin{pmatrix}
\mathcal{L}({\bf r}) & \mathcal{M}({\bf r}) \\ -\mathcal{M}^*({\bf r}) &
-\mathcal{L}({\bf r})
\end{pmatrix}
\begin{pmatrix}
u_q({\bf r}) \\ v_q({\bf r})
\end{pmatrix}
=\hbar \omega_q
\begin{pmatrix}
u_q({\bf r}) \\ v_q({\bf r})
\end{pmatrix},
\end{equation}
where $\mathcal{L}({\bf r})=\mathcal{H}-\mu+2g|\psi({\bf  r})|^2$, $\mathcal{M}({\bf
  r})=g \psi^2({\bf r})$, $u_q({\bf r})$ and
$v_q({\bf r})$ are the quasiparticle amplitudes and $\omega_q$ is the eigenfrequency of the mode $q$. From Eqs.~\eqref{statdec}
and~\eqref{vartheta} one notes that the existence of non-real eigenfrequencies $\omega_q$ results in exponential initial growth of the
population of the corresponding mode, {\it i.e.}, the corresponding
  stationary state is dynamically unstable.
Naturally, in the presence of such dynamically
destabilizing modes, Eqs.~\eqref{vartheta} and~\eqref{bogo} hold only for the initial dynamics of the condensate for which the second and
higher powers of~$\vartheta({\bf r},t)$ are negligible.

Since we consider rotationally symmetric external potentials given by Eq.~\eqref{trap}, the minimal-energy stationary state containing a
$\kappa$-quantum vortex can be written as
 \begin{eqnarray}\label{stastate}\psi({\bf  r},t)=\psi(r,z)e^{i \kappa \theta-i\mu t/\hbar},
\end{eqnarray}
where $(r,\theta,z)$ are the cylindrical coordinates. For this kind of stationary state the azimuthal angle dependence can be separated from the
solutions of the Bogoliubov equations~\eqref{bogo}, and hence the quasiparticle amplitudes can be expressed as
\begin{eqnarray}\label{uv}
u_q({\bf r})&=&u_q(r,z)e^{i(\kappa_q+\kappa)\theta}, \\ v_q({\bf r})&=&v_q(r,z)e^{i(\kappa_q-\kappa)\theta},
\end{eqnarray}
where $\kappa_q$ denotes the angular momentum quantum number of the mode. The dynamical stability of a
given stationary $\kappa$-quantum vortex state can thus be determined by
solving effectively two-dimensional Bogoliubov equations for all quantum numbers
$\kappa_q$. In practice, however, there exist complex eigenfrequencies only
for values of $\kappa_q$ in the vicinity of $\kappa$~\cite{Pu1999a,Kawaguchi2004}.

By scaling time with $1/\omega_r$, length with $a_r=\sqrt{\hbar/m \omega_r}$ and energy with $\hbar \omega_r$, one obtains a dimensionless GP
equation. From this equation, it is observed that the stationary $\kappa$-quantum vortex states are completely determined by the trap
asymmetry parameter~$\lambda$ and the effective interaction strength
$\tilde{g}=4\pi a N/a_r$. Thus we will study the dynamical stability of vortex
states as functions of these parameters.

\section{\label{res}Results}
We have solved the stationary states and the corresponding Bogoliubov excitation spectra for two-, \mbox{three-,} and four-quantum vortices in the
parameter ranges $0 < 1/\lambda \le 2$ and $0 \le \tilde{g} \le 1000$. For typical parameter values used in the experiments~\cite{Shin2004a},
the value $\tilde{g}=1000$ of the interaction strength corresponds to the particle number $N \approx 4 \cdot 10^4$. To investigate condensates
with higher particle numbers, we have also calculated the excitation spectra for $\lambda=1$ and $0 \le \tilde{g} \le 5000$.

\subsection{\label{subA}Two-quantum vortex}
\begin{figure}[!h]
\begin{center}
\includegraphics[width=8.0cm]{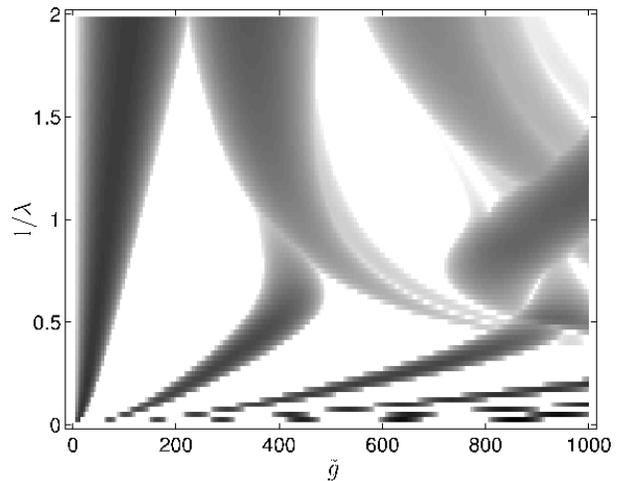}
\caption{\label{double_phasediagram} Gray scale plot of the maximum
imaginary part of the Bogoliubov eigenvalues as a function of the
interaction strength $\tilde{g}=4\pi a N/a_r$ and
  inverse trap anisotropy $1/\lambda=\omega_r/\omega_z$ for the two-quantum vortex state. White denotes the dynamically stable regions with vanishing imaginary parts, and black corresponds to maximum imaginary
part $0.185\omega_r$, the scale in between being linear. The strip-like instability regions for small $1/\lambda$ are in fact continuous,
although they seem fragmentary due to discretization of the parameter space in the computations.}
\end{center}
\end{figure}

\begin{figure}[!h]
\begin{center}
\includegraphics[width=8.0cm]{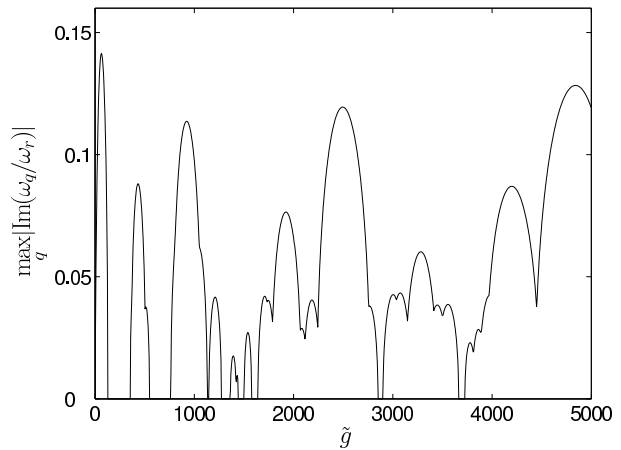}
\caption{\label{double_1d} Maximum imaginary part of the Bogoliubov eigenvalues for the two-quantum vortex state as a function of the
interaction strength $\tilde{g}=4\pi a N/a_r$ for the spherically symmetric trap geometry with $\lambda=1$.}
\end{center}
\end{figure}

Figure~\ref{double_phasediagram} shows the maximum imaginary parts of the eigenfrequencies $\omega_q$ for two-quantum vortex states. In this
case, there is only one angular momentum quantum number, namely $\kappa_q=2$, for which complex frequencies exist. In the limit $1/\lambda
\rightarrow 0$ corresponding to thin pancake shaped condensates, the imaginary part $\textrm{Im}(\omega_q)$ attains a quasi-periodic form as a
function of the interaction strength $\tilde{g}$. In this limit, the results approach those obtained from the effectively one-dimensional
calculations reported in Ref.~\cite{Pu1999a}. As $1/\lambda$ increases towards unity corresponding to the spherical trap shape, the instability
regions widen with respect to the interaction parameter.
However, we note that for the two-quantum vortex state there are fairly large regions in the parameter space where the vortex state is
dynamically stable. Only for slightly prolate trap geometries, $1/\lambda \approx 2$, the instability regions start to overlap, and the vortex
state is dynamically unstable for most interaction strengths~$\tilde{g}$ under consideration. In Fig.~\ref{double_1d}, the maximum imaginary
part of the eigenmode frequencies is shown for interaction strength values in the range $0 \le \tilde{g} \le 5000$, in the special case of the
spherically symmetric trap with $\lambda=1$. The largest dynamical stability regions are seen to exist at fairly weak interaction strengths.

\subsection{\label{subB}Three-quantum vortex}
\begin{figure}[h]
\begin{center}
\includegraphics[width=8.0cm]{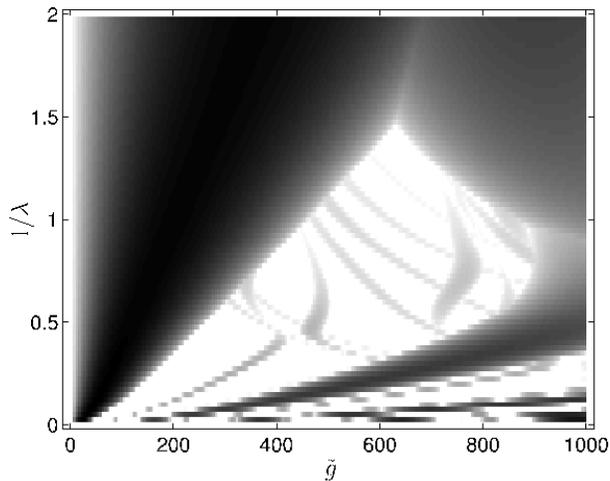}
\caption{\label{triple_phasediagram} Maximum imaginary part of the
  Bogoliubov eigenvalues as a function of the interaction strength $\tilde{g}=4\pi a N/a_r$ and
  inverse trap anisotropy $1/\lambda$ for the three-quantum vortex state.
In the grayscale white corresponds to dynamically stable state and
black the imaginary part value $0.208\omega_r$ of the frequency.
The large instability regions are due to modes
with $\kappa_q=2$, and the thin lines in between them are
caused by modes with $\kappa_q=3$.}
\end{center}
\end{figure}

\begin{figure}[h]
\begin{center}
\includegraphics[width=8.0cm]{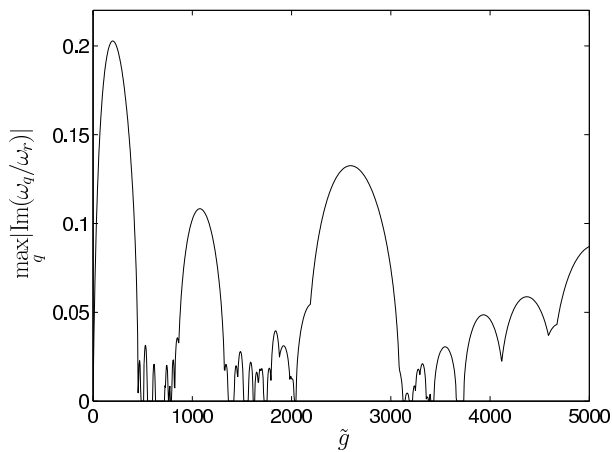}
\caption{\label{triple_1d} Maximum imaginary part of the
  Bogoliubov eigenvalues for the three-quantum vortex state as a function of the interaction strength
  $\tilde{g}=4\pi a N/a_r$ in the spherically symmetric trap geometry.}
\end{center}
\end{figure}

Similarly to the two-quantum vortex case, the maximum imaginary parts of the Bogoliubov eigenfrequencies are shown in
Fig.~\ref{triple_phasediagram} for stationary three-quantum vortex states. In this case, there are three angular momentum quantum numbers,
$\kappa_q=2, 3$, and $4$, which yield complex eigenfrequencies. There exist stable regions also for the three-quantum vortex state, but they are
considerably smaller compared with the two-quantum vortex. The large regions of instability are due to modes with quantum number $\kappa_q=2$,
while the thinner line-like regions originate from modes with $\kappa_q=3$. There exist complex eigenfrequencies also for $\kappa_q=4$, but
their absolute value is significantly smaller compared with the ones for $\kappa_q=2$ and~$3$. The similar shape of the instability regions
corresponding to the quantum number $\kappa_q=2$ for the two- and three-quantum vortex states is also noticeable. Figure~\ref{triple_1d} shows
the maximum imaginary parts of the eigenfrequencies for higher interaction strengths up to $\tilde{g} = 5000$ for three-quantum vortex states in
the spherically symmetric trap geometry.

\subsection{\label{subC}Four-quantum vortex}
\begin{figure}[!h]
\begin{center}
\includegraphics[width=8.0cm]{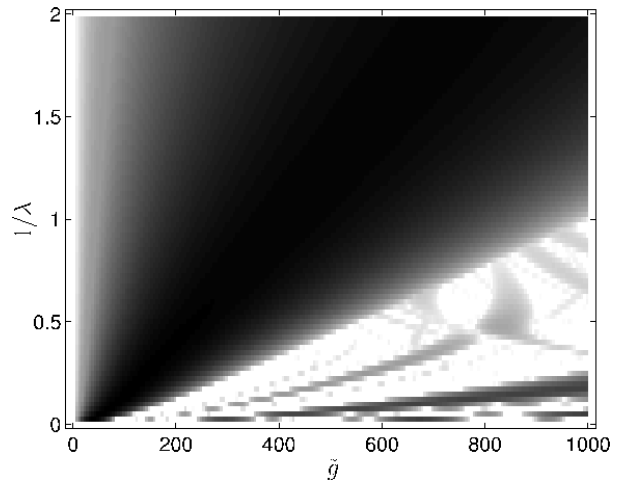}
\caption{\label{quadruple_phasediagram} Maximum imaginary part of the
  Bogoliubov eigenvalues as a function of the interaction strength $\tilde{g}=4\pi a N/a_r$ and
  inverse trap anisotropy $1/\lambda$ for the four-quantum vortex
  state. Complex frequencies exist in this case for
  excitation quantum numbers $\kappa_q=2, 3, 4$, and $5$.
In the linear grayscale black corresponds to imaginary part value
$0.231\omega_r$ of the frequency and the white to zero.
}
\end{center}
\end{figure}

\begin{figure}[!h]
\begin{center}
\includegraphics[width=8.0cm]{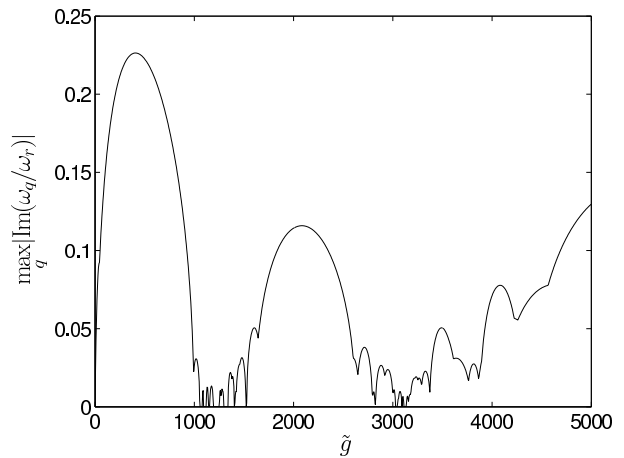}
\caption{\label{quadruple_1d} Maximum imaginary part of the
  Bogoliubov eigenvalues for the four-quantum vortex state as a function of
  the interaction strength
  $\tilde{g}=4\pi a N/a_r$ in the spherically symmetric trap geometry.}
\end{center}
\end{figure}

Figure~\ref{quadruple_phasediagram} shows the maximum imaginary parts of the Bogoliubov eigenfrequencies for the four-quantum vortex state. In
this case, complex eigenfrequencies exist for excitation quantum numbers $\kappa_q=2, 3, 4$, and $5$. As for the two- and three-quantum vortex
states, modes corresponding to the excitation quantum number $\kappa_q=2$ are again dominating, {\it i.e.}, have the largest imaginary parts,
for most of the parameter values. However, in the limit of weak interaction strength, $\tilde{g} \rightarrow 0$, a mode with $\kappa_q=3$
dominates the dynamical instability of the vortex state. Figure~\ref{quadruple_1d} shows the maximum imaginary parts for $\lambda=1$ and $0 \le
\tilde{g} \le 5000$. There are only a few interaction strength values for which the four-quantum vortex is dynamically stable in this geometry.

Comparison of Figs.~\ref{triple_1d} and~\ref{quadruple_1d} shows that the dynamical instability is suppressed for both the three and four
quantum vortex states for $\tilde{g}\approx 3100$. Actually, they are both dynamically stable for $\tilde{g}=3135$. This stability window can
possibly be utilized to stabilize the coexisting three- and four-quantum vortices created in the topological phase imprinting method for $F=2$
condensates.

\section{\label{con}Conclusions}

We have studied the dynamical stability of two-, \mbox{three-,} and four-quantum vortex states in dilute BECs confined in rotationally symmetric traps.
Experimentally, multiquantum vortex states have primarily been realized in long cigar-shaped traps, in which they are known to be dynamically
unstable and to split rapidly into single-quantum vortices. By calculating the Bogoliubov excitation spectra of vortex states as functions of
the trap asymmetry parameter and the condensate particle number, we have shown that there exists regions in which the splitting instabilities
are suppressed and the multiquantum vortices are dynamically stable. Especially, the stability regions of the two-quantum vortex states are
large, and extend from oblate pancake-shaped condensates beyond the spherical geometry. For increasing vorticity quantum number the number of
destabilizing modes grows and the stability regions shrink considerably, but even the four-quantum vortex state can be made dynamically stable
in the spherical geometry for certain condensate particle numbers. We assume that the  predicted suppression of multiquantum vortex splitting
instabilities for certain values of trap asymmetry parameter and particle number can be experimentally verified.

\begin{acknowledgements}
The authors thank CSC, the Finnish IT Center for Science, for computational
resources. J. H. and M. M. acknowledge the Finnish Cultural Foundation for
financial support. Y.\ Shin is appreciated for helpful discussion concerning
the experiments.
\end{acknowledgements}

\bibliography{manu_dynamical_insta}

\begin{thebibliography}{14}
\expandafter\ifx\csname natexlab\endcsname\relax\def\natexlab#1{#1}\fi
\expandafter\ifx\csname bibnamefont\endcsname\relax
  \def\bibnamefont#1{#1}\fi
\expandafter\ifx\csname bibfnamefont\endcsname\relax
  \def\bibfnamefont#1{#1}\fi
\expandafter\ifx\csname citenamefont\endcsname\relax
  \def\citenamefont#1{#1}\fi
\expandafter\ifx\csname url\endcsname\relax
  \def\url#1{\texttt{#1}}\fi
\expandafter\ifx\csname urlprefix\endcsname\relax\def\urlprefix{URL }\fi
\providecommand{\bibinfo}[2]{#2}
\providecommand{\eprint}[2][]{\url{#2}}

\bibitem[{\citenamefont{Matthews et~al.}(1999)\citenamefont{Matthews, Anderson,
  Haljan, Hall, Wieman, and Cornell}}]{Matthews1999b}
\bibinfo{author}{\bibfnamefont{M.~R.} \bibnamefont{Matthews}},
  \bibinfo{author}{\bibfnamefont{B.~P.} \bibnamefont{Anderson}},
  \bibinfo{author}{\bibfnamefont{P.~C.} \bibnamefont{Haljan}},
  \bibinfo{author}{\bibfnamefont{D.~S.} \bibnamefont{Hall}},
  \bibinfo{author}{\bibfnamefont{C.~E.} \bibnamefont{Wieman}},
  \bibnamefont{and} \bibinfo{author}{\bibfnamefont{E.~A.}
  \bibnamefont{Cornell}}, \bibinfo{journal}{Phys. Rev. Lett.}
  \textbf{\bibinfo{volume}{83}}, \bibinfo{pages}{2498} (\bibinfo{year}{1999}).

\bibitem[{\citenamefont{Fetter and Svidzinsky}(2001)}]{Fetter2001c}
\bibinfo{author}{\bibfnamefont{A.~L.} \bibnamefont{Fetter}} \bibnamefont{and}
  \bibinfo{author}{\bibfnamefont{A.~A.} \bibnamefont{Svidzinsky}},
  \bibinfo{journal}{J. Phys.: Condens. Matter} \textbf{\bibinfo{volume}{13}},
  \bibinfo{pages}{R135} (\bibinfo{year}{2001}).

\bibitem[{\citenamefont{Leanhardt et~al.}(2002)\citenamefont{Leanhardt,
  G{\"o}rlitz, Chikkatur, Kielpinski, Shin, Pritchard, and
  Ketterle}}]{Leanhardt2002a}
\bibinfo{author}{\bibfnamefont{A.~E.} \bibnamefont{Leanhardt}},
  \bibinfo{author}{\bibfnamefont{A.}~\bibnamefont{G{\"o}rlitz}},
  \bibinfo{author}{\bibfnamefont{A.~P.} \bibnamefont{Chikkatur}},
  \bibinfo{author}{\bibfnamefont{D.}~\bibnamefont{Kielpinski}},
  \bibinfo{author}{\bibfnamefont{Y.}~\bibnamefont{Shin}},
  \bibinfo{author}{\bibfnamefont{D.~E.} \bibnamefont{Pritchard}},
  \bibnamefont{and} \bibinfo{author}{\bibfnamefont{W.}~\bibnamefont{Ketterle}},
  \bibinfo{journal}{Phys. Rev. Lett.} \textbf{\bibinfo{volume}{89}},
  \bibinfo{pages}{190403} (\bibinfo{year}{2002}).

\bibitem[{\citenamefont{Nakahara et~al.}(2000)\citenamefont{Nakahara, Isoshima,
  Machida, Ogawa, and Ohmi}}]{Nakahara2000a}
\bibinfo{author}{\bibfnamefont{M.}~\bibnamefont{Nakahara}},
  \bibinfo{author}{\bibfnamefont{T.}~\bibnamefont{Isoshima}},
  \bibinfo{author}{\bibfnamefont{K.}~\bibnamefont{Machida}},
  \bibinfo{author}{\bibfnamefont{S.-I.} \bibnamefont{Ogawa}}, \bibnamefont{and}
  \bibinfo{author}{\bibfnamefont{T.}~\bibnamefont{Ohmi}},
  \bibinfo{journal}{Physica B} \textbf{\bibinfo{volume}{284--288}},
  \bibinfo{pages}{17} (\bibinfo{year}{2000}).

\bibitem[{\citenamefont{Isoshima et~al.}(2000)\citenamefont{Isoshima, Nakahara,
  Ohmi, and Machida}}]{Isoshima2000a}
\bibinfo{author}{\bibfnamefont{T.}~\bibnamefont{Isoshima}},
  \bibinfo{author}{\bibfnamefont{M.}~\bibnamefont{Nakahara}},
  \bibinfo{author}{\bibfnamefont{T.}~\bibnamefont{Ohmi}}, \bibnamefont{and}
  \bibinfo{author}{\bibfnamefont{K.}~\bibnamefont{Machida}},
  \bibinfo{journal}{Phys. Rev. A} \textbf{\bibinfo{volume}{61}},
  \bibinfo{pages}{063610} (\bibinfo{year}{2000}).

\bibitem[{\citenamefont{Ogawa et~al.}(2002)\citenamefont{Ogawa, M\"ott\"onen,
  Nakahara, Ohmi, and Shimada}}]{Ogawa2002}
\bibinfo{author}{\bibfnamefont{S.-I.} \bibnamefont{Ogawa}},
  \bibinfo{author}{\bibfnamefont{M.}~\bibnamefont{M\"ott\"onen}},
  \bibinfo{author}{\bibfnamefont{M.}~\bibnamefont{Nakahara}},
  \bibinfo{author}{\bibfnamefont{T.}~\bibnamefont{Ohmi}}, \bibnamefont{and}
  \bibinfo{author}{\bibfnamefont{H.}~\bibnamefont{Shimada}},
  \bibinfo{journal}{Phys. Rev. A} \textbf{\bibinfo{volume}{66}},
  \bibinfo{pages}{013617} (\bibinfo{year}{2002}).

\bibitem[{\citenamefont{M\"ott\"onen et~al.}(2002)\citenamefont{M\"ott\"onen,
  Matsumoto, Nakahara, and Ohmi}}]{Mottonen2002}
\bibinfo{author}{\bibfnamefont{M.}~\bibnamefont{M\"ott\"onen}},
  \bibinfo{author}{\bibfnamefont{N.}~\bibnamefont{Matsumoto}},
  \bibinfo{author}{\bibfnamefont{M.}~\bibnamefont{Nakahara}}, \bibnamefont{and}
  \bibinfo{author}{\bibfnamefont{T.}~\bibnamefont{Ohmi}}, \bibinfo{journal}{J.
  Phys.: Condens. Matter} \textbf{\bibinfo{volume}{14}}, \bibinfo{pages}{13481}
  (\bibinfo{year}{2002}).

\bibitem[{\citenamefont{Leanhardt et~al.}(2003)\citenamefont{Leanhardt, Shin,
  Kielpinski, Pritchard, and Ketterle}}]{Leanhardt2003a}
\bibinfo{author}{\bibfnamefont{A.~E.} \bibnamefont{Leanhardt}},
  \bibinfo{author}{\bibfnamefont{Y.}~\bibnamefont{Shin}},
  \bibinfo{author}{\bibfnamefont{D.}~\bibnamefont{Kielpinski}},
  \bibinfo{author}{\bibfnamefont{D.~E.} \bibnamefont{Pritchard}},
  \bibnamefont{and} \bibinfo{author}{\bibfnamefont{W.}~\bibnamefont{Ketterle}},
  \bibinfo{journal}{Phys. Rev. Lett.} \textbf{\bibinfo{volume}{90}},
  \bibinfo{pages}{140403} (\bibinfo{year}{2003}).

\bibitem[{\citenamefont{{M{\" o}tt{\" o}nen} et~al.}(2003)\citenamefont{{M{\"
  o}tt{\" o}nen}, {Mizushima}, {Isoshima}, {Salomaa}, and
  {Machida}}}]{Mottonen2003a}
\bibinfo{author}{\bibfnamefont{M.}~\bibnamefont{{M{\" o}tt{\" o}nen}}},
  \bibinfo{author}{\bibfnamefont{T.}~\bibnamefont{{Mizushima}}},
  \bibinfo{author}{\bibfnamefont{T.}~\bibnamefont{{Isoshima}}},
  \bibinfo{author}{\bibfnamefont{M.~M.} \bibnamefont{{Salomaa}}},
  \bibnamefont{and}
  \bibinfo{author}{\bibfnamefont{K.}~\bibnamefont{{Machida}}},
  \bibinfo{journal}{Phys. Rev. A} \textbf{\bibinfo{volume}{68}},
  \bibinfo{pages}{023611} (\bibinfo{year}{2003}).

\bibitem[{\citenamefont{Huhtam{\" a}ki et~al.}(2006)\citenamefont{Huhtam{\"
  a}ki, {M{\" o}tt{\" o}nen}, {Isoshima}, Pietil{\" a}, and
  Virtanen}}]{Huhtamaki2006a}
\bibinfo{author}{\bibfnamefont{J.~A.~M.} \bibnamefont{Huhtam{\" a}ki}},
  \bibinfo{author}{\bibfnamefont{M.}~\bibnamefont{{M{\" o}tt{\" o}nen}}},
  \bibinfo{author}{\bibfnamefont{T.}~\bibnamefont{{Isoshima}}},
  \bibinfo{author}{\bibfnamefont{V.}~\bibnamefont{Pietil{\" a}}},
  \bibnamefont{and} \bibinfo{author}{\bibfnamefont{S.~M.~M.}
  \bibnamefont{Virtanen}}, \bibinfo{journal}{quant-ph/0605125}
  (\bibinfo{year}{2006}).

\bibitem[{\citenamefont{Shin et~al.}(2004)\citenamefont{Shin, Saba,
  Vengalattore, Pasquini, Sanner, Leanhardt, Prentiss, Pritchard, and
  Ketterle}}]{Shin2004a}
\bibinfo{author}{\bibfnamefont{Y.}~\bibnamefont{Shin}},
  \bibinfo{author}{\bibfnamefont{M.}~\bibnamefont{Saba}},
  \bibinfo{author}{\bibfnamefont{M.}~\bibnamefont{Vengalattore}},
  \bibinfo{author}{\bibfnamefont{T.}~\bibnamefont{Pasquini}},
  \bibinfo{author}{\bibfnamefont{C.}~\bibnamefont{Sanner}},
  \bibinfo{author}{\bibfnamefont{A.}~\bibnamefont{Leanhardt}},
  \bibinfo{author}{\bibfnamefont{M.}~\bibnamefont{Prentiss}},
  \bibinfo{author}{\bibfnamefont{D.}~\bibnamefont{Pritchard}},
  \bibnamefont{and} \bibinfo{author}{\bibfnamefont{W.}~\bibnamefont{Ketterle}},
  \bibinfo{journal}{Phys. Rev. Lett.} \textbf{\bibinfo{volume}{93}},
  \bibinfo{pages}{160406} (\bibinfo{year}{2004}).

\bibitem[{\citenamefont{Kumakura et~al.}(2006)\citenamefont{Kumakura, Hirotani,
  Okano, Yabuzaki, and Takahashi}}]{Kumakura2006a}
\bibinfo{author}{\bibfnamefont{M.}~\bibnamefont{Kumakura}},
  \bibinfo{author}{\bibfnamefont{T.}~\bibnamefont{Hirotani}},
  \bibinfo{author}{\bibfnamefont{M.}~\bibnamefont{Okano}},
  \bibinfo{author}{\bibfnamefont{T.}~\bibnamefont{Yabuzaki}}, \bibnamefont{and}
  \bibinfo{author}{\bibfnamefont{Y.}~\bibnamefont{Takahashi}},
  \bibinfo{journal}{Phys. Rev. A} \textbf{\bibinfo{volume}{73}},
  \bibinfo{pages}{063605} (\bibinfo{year}{2006}).

\bibitem[{\citenamefont{Pu et~al.}(1999)\citenamefont{Pu, Law, Eberly, and
  Bigelow}}]{Pu1999a}
\bibinfo{author}{\bibfnamefont{H.}~\bibnamefont{Pu}},
  \bibinfo{author}{\bibfnamefont{C.~K.} \bibnamefont{Law}},
  \bibinfo{author}{\bibfnamefont{J.~H.} \bibnamefont{Eberly}},
  \bibnamefont{and} \bibinfo{author}{\bibfnamefont{N.~P.}
  \bibnamefont{Bigelow}}, \bibinfo{journal}{Phys. Rev. A}
  \textbf{\bibinfo{volume}{59}}, \bibinfo{pages}{1533} (\bibinfo{year}{1999}).

\bibitem[{\citenamefont{Kawaguchi and Ohmi}(2004)}]{Kawaguchi2004}
\bibinfo{author}{\bibfnamefont{Y.}~\bibnamefont{Kawaguchi}} \bibnamefont{and}
  \bibinfo{author}{\bibfnamefont{T.}~\bibnamefont{Ohmi}},
  \bibinfo{journal}{Phys. Rev. A} \textbf{\bibinfo{volume}{70}},
  \bibinfo{pages}{043610} (\bibinfo{year}{2004}).

\end{thebibliography}

\end{document}